\documentclass[twocolumn,showpacs,preprintnumbers,amsmath,amssymb]{revtex4}

\usepackage{graphicx}
\usepackage{dcolumn}
\usepackage{bm}

\newcommand{\beq}{\begin{eqnarray}}
\newcommand{\eeq}{\end{eqnarray}}
\newcommand{\nn}{\nonumber}
\newcommand{\lanln}[3]{\texttt{arXiv:#1#2\ [#3]}}   

\begin{document}


\title{Ostrogradski approach for the Regge--Teitelboim 
type cosmology}

\author{Rub\'en Cordero}
\email{cordero@esfm.ipn.mx}
\affiliation
{
Departamento de F\'\i sica, Escuela Superior de F\'\i sica y Matem\'aticas del
IPN,    \\ 
Unidad Adolfo L\'opez Mateos, Edificio 9, 07738 M\'exico, Distrito Federal, M\'exico
}
\author{Alberto Molgado}
\email{amolgado@planck.reduaz.mx}
\affiliation{Unidad Acad\'emica de F\'\i sica, Universidad Aut\'onoma de
Zacatecas,         \\
Calzada Solidaridad esq.~Paseo a la Bufa s/n, 98062 Zacatecas, Zacatecas, M\'exico}
\author{Efra\'\i n Rojas}
 \email{efrojas@uv.mx}
\affiliation{Departamento de F\'\i sica, Facultad de F\'\i sica e Inteligencia
Artificial,    \\ Universidad Veracruzana, 91000 Xalapa, Veracruz, M\'exico}

\date{\today}

\begin{abstract}
We present an alternative geometric inspired derivation of the quantum cosmology 
arising from a brane universe in the context of {\it geodetic gravity}. 
We set up the Regge--Teitelboim model to describe our universe, and we 
recover its original dynamics by thinking of such field theory 
as a second--order derivative theory. We 
refer to an Ostrogradski Hamiltonian formalism to prepare the system 
to its quantization. Our analysis highlights the 
second-order derivative nature of the RT model and the inherited geometrical 
aspect of the theory. A canonical transformation brings us to the internal 
physical geometry of the theory and induces its quantization straightforwardly.
By using the Dirac canonical quantization method our approach comprises 
the management of both first- and second-class constraints where the counting of 
degrees of freedom follows accordingly. At the quantum level our Wheeler--De 
Witt equation agrees with previous results recently found.
On these lines, we also comment upon the compatibility of our approach with the  
Hamiltonian approach proposed by Davidson and coworkers.
\end{abstract}

\pacs{04.50.-h, 04.60.Kz, 04.60.Ds, 98.80.Jk, 98.80.Qc}

\maketitle

\section{\label{sec:intro}Introduction}

The concept of a relativistic extended object as a surface immersed
in a bulk has increased the interest in physics due to
its wide range of applications. One can model, for example, the smallest 
physical entities, like quarks, as vibrations of strings up to the 
entire universe as a relativistic extended object.  
Along a related line, with the advent of brane world universes, cosmology 
in the presence of extra dimensions has been the subject of intense research.
In fact, the idea that our universe could be a $3+1$ dimensional 
surface embedded in a higher dimensional spacetime was set up by 
Regge and Teiltelboim (RT) a long time ago~\cite{RT} and pursued by many 
authors~\cite{tapia,pavsic,pavsic1,maia}. The scope of such a model is 
that gravitation can be described in a point- or stringlike fashion, 
as the worldvolume swept out by  the motion of a three-dimensional spacelike 
brane evolving in a higher-dimensional bulk spacetime~\cite{RT}.
Recently, the RT brane model has been considered as one of the two main
pillars of a unified branelike theory~\cite{DG}, where the Randall-Sundrum
brane theory \cite{Randall} is included. When one addresses this issue in a Minkowski spacetime, the 
model is named {\it geodetic gravity}, and it has been extensively 
studied by Davidson and coworkers~\cite{davidson0,davidson,DKL,davidson1}. 
Although the RT model is not the most popular theory for brane world 
universes~(at the end of last century, there was a revival of the idea that our 
universe could be a hypersurface; see, for example,~\cite{Randall,Arkani}),
it is very stimulating while thinking in the spirit of brane gravity {\it \`{a} la} string.
The cosmology that arises from this model is interesting in its 
own right since it provides an alternative route to better understand 
classical cosmology in extra dimensions, and also it supplies a compelling
model to apply the canonical quantization methods. Indeed, in the context 
of quantum brane cosmology~\cite{SSP} our universe can be explained 
through a tunneling process where the well-known problem of boundary 
conditions of four-dimensional cosmology is solved \cite{Alex,CV}.

In most field theories the action depends usually on the fields and their 
first derivatives. By contrast, the RT model is a genuine second-order 
derivative model in the field variables, which are the embedding functions 
rather than the induced metric. Generally, one identifies and neglects 
a surface term associated with the linear dependence of the accelerations. 
Similarly as in general relativity, it is 
a well-known fact that a ``harmless'' surface term can be neglected or removed
at the beginning as occurs with the well-known Gibbons-Hawking-York term into 
the action. Whichever field context, the extremization of the corresponding action yields equations
of motion of second order in derivatives in the field variables. 
Thus the RT model, like the Einstein-Hilbert action for general relativity, becomes 
transformed in an effective first-order field theory. However, by what formally 
appears to be a customary procedure, to follow such a strategy raises important 
limitations especially in the Hamiltonian framework for the RT model where it leads 
to certain troubles, 
as was noticed first by Regge and Teitelboim, due mainly to the fact that 
the scalar constraint is not written down in a closed form straightforwardly. 
In pursuing this endeavor, Davidson and coworkers tackled the problem successfully. 
They considered an extra nondynamical canonical field $\lambda$ in the first-order 
Hamiltonian framework in order to get quadratic constraints of the phase 
space that recuperate the dynamics accordingly \cite{davidson0,davidson1,DKL}. 
The explicit handling of the quantum RT model is made possible by extending the 
ordinary phase space, which in turn provides a wealth of information of the 
cosmology that this model possesses.

In the present paper we consider an alternative formulation for geodetic 
gravity which is strongly based in the Ostrogradski program for higher-order 
derivative theories~\cite{Ostro, nesterenko}.  For second-order theories this 
approach treats the velocities as independent 
fields. This is an unconventional viewpoint for the RT theory, and one might 
therefore wonder if such a description is viable at all since this does not 
necessarilly represent a shortcoming: for this special case the
addition of more degrees of freedom is physically more accurate, but it means
then that the first-order theory is incomplete in some sense. For this reason, 
it seems promising to start directly from the full RT model instead of
omitting the surface term {\it a priori}.
Hence, we pay close attention to a Hamiltonian approach for geodetic gravity
constructed  by appliying the Ostrogradski scheme which in turn leads 
to the correct dynamics. In particular, it is of a great interest to use 
the full model straightforwardly 
for obtaining the quantum approach for brane cosmology. 
Our intention is to cope directly with the inherent second-order derivative nature of the RT model. 
As discussed below, we gain certain improvements of clarity 
by the use of this formalism in comparison with previous works. 
Contrary to the standard quadratic form of the constraints
for ordinary first--order reparametrization invariant
theories, in the Ostrogradski approach for RT field
theory the constraints are projections of the momenta 
along the velocities as well as along the unit spacelike normal vector to the 
brane. To illustrate our development we specialize our considerations to a 
minisuperspace model where the inherent gauge invariance under the reparametrization 
of time is evident.
We show that the canonical Dirac constraint quantization of this model casts into
a satisfactory Wheeler--De Witt (WDW) equation on the wave function for a  branelike 
universe. The handling of the quantum approach is made possible by a canonical 
transformation which results to be a Lorentz rotation in phase space. Such a transformation 
brings our constraints into a physically meaningful set which enable us to follow the standard 
Dirac constraint quantization programme.
Our quantum treatment hence leads to a well-defined Wheeler-DeWitt 
equation which, even though it is technically complicated to solve,
presents the right behavior for the quantum potential, 
estimating the accuracy of our approach.

The outline of the paper is as follows. In Sec. II we briefly review 
some geometrical aspects of the RT model for a general $d$-dimensional 
brane, which are of interest for the rest of the paper. This section will 
serve to explain our notation and to gain insight into our Hamiltonian 
approach. In Sec. III we adapt our approach to a minisuperspace model 
in which we specialize to the geometry generated by the Friedman-Robertson-Walker 
(FRW) model. We explicitly give the Lagrangian density associated to the RT model 
which includes the surface term. Sec. IV deals with the Ostrogradski 
approach for the model we are considering, and we develop the corresponding 
constraint analysis. In Sec. V we propose the gauge-fixing for the model 
in order to completely identify the structure of the reduced phase space.
In Sec. VI, we study the quantization of our model within the scheme of 
Dirac quantization.  Finally, in Sec. VII we draw some conclusions. As a 
general feature, our presentation avoids cumbersome notation and is intended to 
be index-free as possible.



\section{Regge--Teitelboim model}

Consider a brane $\Sigma$ of dimension $d$, evolving in a fixed 
Minkowski $N$ dimensional background spacetime with metric 
$\eta_{\mu \nu}$. Its trajectory, or worldvolume $m$ of dimension 
$d+1$, is described by the embedding $x^{\mu}=X^{\mu}(\xi^{a})$, 
where $x^{\mu}$ are local coordinates for the background spacetime, 
$\xi^{a}$  local coordinates for $m$, and $X^\mu$ 
the embedding functions ($\mu ,\nu= 0,1, \ldots, N-1$; $a,b=0,1,
\ldots,d$). We denote by $e^\mu{}_a = \partial_a X^\mu$ the tangent
vectors to $m$. 
In this framework we introduce $N-d-1$ unit normal vectors to the 
worldvolume, denoted by $n^\mu {}_i \,\,(i=1,2,\ldots,N-d-1)$. These 
are defined implicitly by $n^i \cdot e_a = 0$, and we choose to normalize 
them as $n_i \cdot n_j = \delta_{ij}$. 

The RT model for a $d$-dimensional brane $\Sigma$ is defined by 
the action functional
\begin{equation}
S _{RT} [X]= \frac{\alpha}{2} \int_m d^{d+1}\xi \,\sqrt{-g}\, 
{\cal R} - \int_m d^{d+1}\xi \,\sqrt{-g}\,\Lambda \,,
\label{eq:action1}
\end{equation}
where the constant $\alpha$ has dimensions $ [L]^{(1-d)}$, $g$ 
denotes the determinant of the induced metric $g_{ab}= \eta_{\mu\nu} \, 
e^\mu {}_a e^\nu{}_b = e_a \cdot e_b$. We have also
included in this action a cosmological constant term, $\Lambda$. 
The extrinsic curvature of $m$ is $K_{ab}{}^i = - n^i \cdot D_a e_b$, where 
$D_a = e^\mu {}_a D_\mu$ and $D_\mu$ is the covariant derivative in the 
bulk spacetime. The mean extrinsic curvature is given by the trace 
$K^i = g^{ab} K_{ab}{}^i$ where $g^{ab}$ denotes the inverse of $g_{ab}$. 
The scalar curvature ${\cal R}$ of $m$ can be obtained either directly from 
the induced metric $g_{ab}$, or, in terms of the extrinsic curvature, 
via the contracted Gauss--Codazzi equation, ${\cal R}= K^{i}K_i - 
K_{ab}^{i}K^{ab}_i$ \cite{defo,spivak}. 

The response of the action (\ref{eq:action1}) to a deformation
of the surface $X \to X + \delta X$ is characterized by a conserved
stress tensor \cite{noether}
\begin{equation}
f^{a\,\mu} = - \left( \alpha {\cal G}^{ab} + \Lambda g^{ab}\right)
e^\mu {}_b  \,,
\label{eq:stress}
\end{equation}
where ${\cal G}_{ab} = {\cal R}_{ab} - \frac{1}{2} {\cal R} g_{ab} $ is 
the worldvolume Einstein tensor, with ${\cal R}_{ab}$ being the Ricci 
tensor. This quantity will provide relevant physical information 
especially with consistent conservation laws.
Following the line of reasoning of \cite{noether}, the classical brane 
trajectories can be obtained from the covariant conservation law, 
$\nabla_a f^{a\,\mu}=0$, where $\nabla_a$ is the covariant derivative
compatible with the induced metric $g_{ab}$ \cite{hamRT}. This yields
\cite{Carter}  
\begin{equation}
T^{ab} \; K_{ab}{}^i = 0\,,
\label{eq:eom}
\end{equation}
where $T^{ab} = \alpha\,{\cal G}^{ab} + \Lambda g^{ab}$. In fact, 
$T^{ab}$ corresponds to the intrinsic stress tensor defined in the 
usual way by $-2/\sqrt{-g}(\delta S_{RT}/\delta g_{ab})$. Its conservation
is supported by the Bianchy identity. The equations of motion (\ref{eq:eom}) 
are of second order in derivatives of the embedding functions because 
of the presence of the extrinsic curvature. This is so even though in 
the scalar curvature ${\cal R}$ we have the presence of the extrinsic 
curvature. Owing to the reparametrization invariance of the RT model, 
there are only $D-d-1$ independent equations, along the normals; the 
remaining $d+1$ tangential components are satisfied identically, as a 
consequence of the reparametrization invariance of the action (\ref{eq:action1}). 

An important quantity constructed with the conserved stress tensor is
given by
\begin{equation}
 \pi^\mu = \eta_a f^{a\,\mu} =  - \left( \alpha 
{\cal G}^{ab} + \Lambda g^{ab}\right)\eta_a e^\mu {}_b 
\label{eq:pi}
\end{equation}
where $\eta^a$ stands for the timelike unit normal vector to the
brane $\Sigma$ when it is viewed into $m$ \cite{defoedges}. In fact, 
Eq. (\ref{eq:pi}) is nothing but the conserved linear momentum associated with
the Noether charge of the action (\ref{eq:action1}) specialized to 
background translations \cite{noether}. The $\Sigma$ basis, $ \left\lbrace 
\epsilon^\mu {}_A, \eta^\mu, n^\mu {}_i \right\rbrace $ satisfies the 
completeness relation
\begin{equation}
\eta^{\mu \nu} = n^{\mu\,i} n^\nu{}_i -\eta^\mu \eta^\nu +
h^{AB} \epsilon^\mu {}_{A}\epsilon^\nu {}_B\,,
\label{eq:Hmunu} 
\end{equation}
where $h_{AB} = g_{ab} \epsilon^a{}_A \epsilon^b{}_B$ is the spatial
metric on $\Sigma$  and $\epsilon^a{}_A$ are the tangent vectors to 
$\Sigma$, \quad ($A,B = 1,2,\ldots,d$). The vector $\eta^\mu$ stands for a timelike
unit vector to $\Sigma$ (see Refs.~\cite{hamRT,hambranes} for more details).

In presence of other possible matter sources with stress tensor $T^{ab} _m =
(-2/\sqrt{-g}) \delta S_m/\delta g_{ab}$, where $S_m$ is a matter action,
we do not expect considerable modifications in our approach. The equations
of motion (\ref{eq:eom}) remain unchanged
in form. It is sufficient to add the matter stress tensor to the original
one described in (\ref{eq:eom}). Similarly, the conserved linear momentum
(\ref{eq:pi}) is unaffected in form when another type of matter is included.
It gets an additional contribution of the form $\pi^\mu _m = -T^{ab} _m \eta_a e^\mu {}_b$.
These nice features allow us to develop straightforwardly a Hamiltonian analysis without 
substantial changes under the inclusion of matter fields. This fact was also noticed
in \cite{davidson1}.

\section{Minisuperspace model}

We turn now to restrict the RT model itself (\ref{eq:action1}) to the case of a 
minisuperspace model. Consider a $3$-brane $\Sigma$, evolving in a 5-dimensional 
Minkowski spacetime, $ds^2 = - dt ^2 + da^2 + a^2 \, d \Omega_3 ^2$, where 
$d \Omega_3 ^2$ stands for the metric of a unit 3-sphere, i.e., $d \Omega_3 ^2 =
d\chi^2 + \sin^2 \chi d\theta^2 + \sin^2 \chi \sin^2 \theta d\phi^2$. For the 
sake of simplicity, we choose the function $ \sin^2 \chi $  in $d \Omega_3 ^2$ 
to consider a closed universe. If
\begin{equation}
x^\mu = X^\mu (\xi^a) 
= \left( t(\tau), a(\tau) , \chi, \theta, \phi \right) 
\end{equation}
is a parametric representation of the trajectory of $\Sigma$, we assure that the 
geometry of the worldvolume generated is that of the FRW case. According to the 
cosmology jargon, $a(\tau)$ is known as the {\it scale factor}.

The basis adapted to the worldvolume is given by the four tangent vectors $e^\mu{}_a 
\quad (a=0,1,2,3)$ together with the unit spacelike normal vector
\begin{equation}
n_\mu = \frac{1}{N}\left( -\dot{a} , \dot{t} , 0, 0, 0 \right) \,,
\end{equation}
where the dot stands for derivation with respect to $\tau$. For short in the 
notation we have introduced the quantity, $N = \sqrt{\dot{t}^2 - \dot{a}^2}$, which 
coincides with the lapse function when we perform an ADM decomposition of the 
action (\ref{eq:action1}) \cite{hambranes,hamRT}.

The metric induced on the worldvolume is given by
\begin{equation}
ds^2 = g_{ab}d\xi^a d\xi^b = -N^2d\tau^2 + a^2 d \Omega_3 ^2 \,.
\label{eq:metric}
\end{equation}
The spatial components of this metric correspond to the metric associated to 
$\Sigma$ when this is described by its embedding in the worldvolume itself. 
Furthermore, for this latter parametrization, we have $\eta^a = 1/N (1,0,0,0)$ 
such that $g_{ab}\epsilon^a{}_A \eta^b = 0$. 

The Ricci scalar associated with the metric (\ref{eq:metric}) reads
\begin{equation}
{\cal R} = \frac{6\dot{t}}{a^2 N^4}\left( a \ddot{a} \dot{t} 
- a \dot{a} \ddot{t} + N^2 \dot{t} \right)\,. 
\end{equation}
The linear dependence that the Ricci scalar possesses in the accelerations 
of the variables $t(\tau)$ and $a(\tau)$ is particularly remarkable.

The Lagrangian density ${\cal L}= \sqrt{-g}( \frac{\alpha}{2} {\cal R} - 
\Lambda) $ thus becomes
\begin{equation}
{\cal L} = \frac{\Upsilon \,a\,\dot{t}\alpha }{N^3}\, \left( a 
\ddot{a} \dot{t} - a \dot{a}  \ddot{t} + \dot{t}^3 - \dot{a}^2 
\dot{t} \right) - \frac{\Upsilon N a^3}{3}\,\Lambda \,,
\label{eq:Lag1}
\end{equation}
where $\Upsilon = 3\sin \theta \sin^2\chi$. Thus, the RT action specialized to 
spherical configurations, in terms of an arbitrary parameter $\tau$, is reduced to 
\begin{equation}
S_{RT} = 6\pi^2 \int d\tau \,L (a,\dot{a},\ddot{a},\dot{t},\ddot{t})\,,
\label{eq:action2}
\end{equation}
where the Lagrangian function is given by~\footnote{Indisputably, 
it can be argued that this Lagrangian function can be brought into the
total derivative of a surface term plus a function depending up to the 
velocities of the variables $a$ and $t$, namely,
$
L = \frac{a\,\dot{t}}{N^3}\, \left( a \ddot{a} \dot{t} - 
a \dot{a}  \ddot{t} + \dot{t}^3 - \dot{a}^2 \dot{t}  \right)
- Na^3H^2 
= - \frac{a\,\dot{a}^2}{N}  + a\,N \left(1- a^2H^2  
\right)  + \frac{d}{d\tau}\left( \frac{a^2\,\dot{a}}{N}\right) \,.
$
In this work we avoid to follow the usual shortcut and we maintain
the surface term which plays an important role in our calculations.
}
\begin{equation}
L (a,\dot{a},\ddot{a},\dot{t},\ddot{t}) = \frac{a\,\dot{t}}{N^3}\, 
\left( a \ddot{a} \dot{t} - a \dot{a}  \ddot{t} + N^2 \dot{t} \right) 
- Na^3 H^2 \,,
\label{eq:Lag2}
\end{equation}
where we have introduced the constant quantity $H^2 := \Lambda/3\alpha$.
Thus, we have only $a(\tau)$ and $t(\tau)$ as independent dynamical variables.
Despite the acceleration dependence in the Lagrangian, as characterizes
second-order derivative theories, the equations of motion remain 
second order in the field variables (see Eq. ~(\ref{eq:eom})).

We proceed now to evaluate both the Einstein and the extrinsic curvature
tensors of the worldvolume as described by the metric~(\ref{eq:metric}). We 
have the nonvanishing explicit components
\begin{eqnarray*}
 {\cal G}^\tau {}_\tau &=& - \frac{3\dot{t}^2}{a^2N^2} \,,
\\
 {\cal G}^\chi {}_\chi ={\cal G}^\theta {}_\theta =  {\cal G}^\phi {}_\phi 
&=& - \frac{\dot{t}^2}{a^2 N^4}\left[ 
2a\dot{t} \frac{d}{d\tau} \left(\frac{\dot{a}}{\dot{t}} \right) + N^2 
\right] \,,
\end{eqnarray*}
and
\begin{eqnarray*}
K^\tau {}_\tau &=& \frac{\dot{t}^2}{N^3}\frac{d}{d\tau}\left(
\frac{\dot{a}}{\dot{t}} \right)\,,
\\
K^\chi {}_\chi =K^\theta {}_\theta = K^\phi {}_\phi &=& 
\frac{\dot{t}}{aN}\,.
\end{eqnarray*}
Clearly, we can read off immediately the spatial components of the extrinsic 
curvature tensor as well as its mean extrinsic curvature given by ${\cal K} 
= h^{AB}K_{AB}= 3\dot{t}/aN$.

As dictated by Eq. ~(\ref{eq:eom}), there is only one equation of motion
given by
\begin{equation}
\frac{d}{d\tau}\left( \frac{\dot{a}}{\dot{t}}\right) = -\frac{N^2}{a\dot{t}} 
\frac{\Theta}{\Phi}\,,
\label{eq:final}
\end{equation}
where we have introduced the functions
\begin{subequations}
 \label{allequations}
\begin{eqnarray}
 \Theta &:= & \dot{t}^2 - 3N^2a^2H^2   \,, \label{equationa} \\
\Phi   &:= & 3\dot{t}^2 - N^2 a^2H^2  \,, \label{equationb}
\end{eqnarray}
\end{subequations}
for simplicity. Eq. ~(\ref{eq:final}) obviously only involves second derivatives of 
the field variables $a$ and $t$. For any solution for $a(\tau)$ we have a gauge freedom 
to choose a function for $t(\tau)$ as we will see below.

\section{Ostrogradski--Hamiltonian approach}

A deeper insight of the phase space structure of the theory defined by the
Lagrangian (\ref{eq:Lag2}) is achieved by an Ostrogradski procedure for
higher-order derivative systems. (A complete description of the Hamiltonian 
formulation for branes whose action depends on the extrinsic curvature of 
their worldvolume is provided in \cite{hambranes,hamRT}.) The highest conjugate 
momenta to the velocities $\left\lbrace \dot{t},\dot{a} \right\rbrace$ are, 
respectively,
\begin{eqnarray}
P_t & = & 
\frac{\partial L}{\partial \ddot{t}} = 
-\frac{a^2 \, \dot{a}\, \dot{t}}{N^3} \,, \\
P_a & = & 
\frac{\partial L}{\partial \ddot{a}} = 
\frac{a^2\, \dot{t}^2}{N^3} \,,
\end{eqnarray}
such that the highest momentum spacetime vector is
\begin{equation}
P_\mu = \frac{a^2 \,\dot{t}}{N^3} \,\left( 
-\dot{a},
\dot{t},
0,
0 , 0
\right)=\frac{a^2 \dot{t}}{N^2}n_\mu\,.
\label{eq:P} 
\end{equation}
Though this momentum has not a direct mechanical meaning it will become 
important to achieve a Legendre transformation in order to obtain the 
Hamiltonian function for our system (see Eq. ~(\ref{eq:ham}) below).
Note that the momentum $P_\mu$ is directed normal to the worldvolume.

The conjugate momenta to the position variables $\left\lbrace t,a \right\rbrace$ 
are, respectively 
\begin{eqnarray}
p_t & = & 
\frac{\partial L}{\partial \dot{t}} - \frac{d}{d\tau} \left( 
\frac{\partial L}{\partial \ddot{t}}\right) \nonumber \\ &=& 
\frac{a\,\dot{t}}{N^3} \left[ \dot{a}^2 + N^2 \left( 1 - a^2 H^2 
\right) \right] =: -\Omega \,,
\label{eq:pt}
\\
p_a & = & 
\frac{\partial L}{\partial \dot{a}} - \frac{d}{d\tau}
\left( \frac{\partial L}{\partial \ddot{a}}\right) \nonumber \\
&=& 
- \frac{a\,\dot{a}}{N^3} \left[ \dot{a}^2 + N^2 \left( 1 - a^2 H^2 
\right)\right] = \left( \frac{\dot{a}}{\dot{t}}\right) \Omega \,. 
\label{eq:pa}
\end{eqnarray}
Important to note is the fact that both momenta, $p_t$ and $p_a$, are 
from a totally different nature.  Indeed, while the momentum $p_t$ is
not influencied at all by the surface terms (as expected),
the momentum $p_a$ is obtained by two contributions:  one coming from the
ordinary theory and the other by a surface term.  In this way, 
we can denote the momentum $p_a$ as 
\beq
\label{eq:pa-terms}
p_a := \mathbf{p}_a + \mathfrak{p}_a \,,
\eeq
where 
\beq
\label{eq:boldp}
\mathbf{p}_a & = & - \frac{a\dot{a}}{N^3}\left[ \dot{a}^2 + N^2 \left( 3 - a^2 H^2 
\right) \right]                          \,,\\
\label{eq:boldq}
\mathfrak{p}_a & = &  \frac{2a\dot{a}}{N}                        \,.
\eeq
It is crucial to recognize them as the canonical momentum worked out in~\cite{DKL} 
and as the momentum conjugated to the $a(\tau)$-variable when considering as the 
Lagrangian only the surface term $L_s = d/d\tau (a^2 a/N)$, respectively.

To see the geometrical structure of this momentum, it is convenient to write
$p_\mu$ as
\begin{eqnarray}
p_\mu  &=&  \frac{\Omega}{\dot{t}} 
\left( -\dot{t}, \dot{a},0,0,0 \right) = \frac{\Omega}{\dot{t}} \dot{X}_\mu  \,.
\label{eq:p} 
\end{eqnarray}
We realize that this momentum is identical with the vector $\pi_\mu$ defined on 
Eq. ~(\ref{eq:pi}), which is the projection of the conserved stress tensor along 
the unit timelike normal vector $\eta^a$ to $\Sigma$. 

The appropiate phase space of the system, $\Gamma := \left\lbrace 
t,a,\dot{t},\-\dot{a};p_t,p_a,P_t,P_a \right\rbrace$,  has been 
identified explicitly. Thus in $\Gamma$, 
the Ostrogradski formalism yields the canonical 
Hamiltonian
\begin{equation}
H_0 = p \cdot \dot{X} + P \cdot \ddot{X} - L = p_a\,\dot{a} + 
p_t\,\dot{t} + J_{{\cal R}}\,,
\label{eq:ham}
\end{equation}
where we have defined
\begin{equation}
J_{{\cal R}} = - \frac{a}{N} \left[ \dot{a}^2 + N^2 \left( 1 - a^2 H^2 \right)\right] =
\frac{N^2}{\dot{t}} \Omega \,.
\end{equation}
This potential-like term results an implicit function of the
phase space variables  in the combination $N^3 P^2$.
At first glance, this may look as an unnecessary complication to write
the phase space quantities in terms of $\Omega$, but this quantity 
results in a physical observable: It is nothing but the conserved
bulk energy. Indeed, squaring the energy equation~(\ref{eq:pt}), 
results in the so--called evolution master  
equation~\cite{RT,pavsic1,tapia,maia} 
\begin{equation}
N^2 + \dot{a}^2 = \gamma\, N^2 a^2 H^2\,,  
\label{eq:master}
\end{equation}
where $\gamma=\gamma(a)$ satisfies the cubic equation $\gamma (\gamma - 1)^2 
= \Omega^2 /a^8 H^6$.

\subsection{Constraint analysis}

Since we are dealing just with a second--order derivative theory
linear in the accelerations, we have two primary constraints given
by the definition of the momenta itself, and hence $\phi_\mu := P_\mu 
- \frac{a^2 \dot{t}}{N^2}n_\mu = 0$. Instead of these constraints, 
here we will follow a different but convenient route. We choose to 
project the momentum~(\ref{eq:P}) along the velocity vector as well 
as the normal vector to $\Sigma$, where in general $n^\mu = n^\mu 
(\dot{X}^\nu)$. This is supported by using the geometrical completeness 
relation (\ref{eq:Hmunu}) in $\phi_\mu = \eta_{\mu \nu}\phi^\nu$. 
Thus, we get the primary constraints
\begin{eqnarray}
{\cal C}_1 &=& P \cdot \dot{X} =0\,, 
\label{eq:c1}\\
{\cal C}_2 &=& N\,P\cdot n
- \frac{a^2\,\dot{t}}{N}=0\,.
\label{eq:c2}
\end{eqnarray}
Therefore, the total Hamiltonian which generates time evolution of the 
fields is
\begin{equation}
\label{eq:Htotal}
H_T = H_0 + \lambda_1\,{\cal C}_1 + \lambda_2\,{\cal C}_2\,, 
\end{equation}
where $\lambda_1$ and $\lambda_2$ are Lagrange multipliers
enforcing the primary constraints.

As customary, time--evolution for any canonical variable $z\in \Gamma$ reads
\begin{equation}
 \dot{z} = \left\lbrace z, H_T \right\rbrace \,, 
\end{equation}
on the constraint surface, where the generalized Poisson bracket for any two 
functions $F(z)$ and $G(z)$ in $\Gamma$ is appropriately defined as
\begin{eqnarray}
\left\lbrace F, G \right\rbrace &=& \frac{\partial F}{\partial t}
\frac{\partial G}{p_t} + \frac{\partial F}{\partial a}
\frac{\partial G}{p_a} + \frac{\partial F}{\partial \dot{t}}
\frac{\partial G}{P_t} + \frac{\partial F}{\partial \dot{a}}
\frac{\partial G}{P_a}  \nonumber \\
&-& ( F \longleftrightarrow G )\,.
\label{eq:PB}
\end{eqnarray}
Important to mention is the fact that the total Hamiltonian~(\ref{eq:Htotal}) 
leads us directly to the right equations of motion~(\ref{eq:final}) through 
the conventional Ostrogradski approach for higher-order derivative 
systems~\cite{nesterenko}.

We also note that under the symplectic structure~(\ref{eq:PB}), the constraints 
(\ref{eq:c1}) and (\ref{eq:c2}) result to be in involution, $\left\lbrace 
{\cal C}_1,{\cal C}_2 \right\rbrace = 0$. According to the Dirac program for 
constrained systems, both ${\cal C}_1$ and ${\cal C}_2$ must be preserved by the 
evolution which demands the existence of the secondary constraints
\begin{eqnarray}
 {\cal C}_3 &=& H_0 = 
p\cdot \dot{X}
+ N \left( a^3 H^2 - \frac{1}{a^3}N^2P^2 \right)\,, 
\label{eq:C3}\\
{\cal C}_4 &= & p\cdot n.
\label{eq:C4}
\end{eqnarray}
The vanishing of the canonical Hamiltonian is expected courtesy of the
reparametrization invariance of the RT model. Hence the canonical Hamiltonian 
$H_0$ generates diffeomorphisms normal to the worldvolume. The secondary 
constraint (\ref{eq:C4}) is characteristic for every brane model linear in 
accelerations. The process of generation of further constraints is stopped 
at this stage since ${\cal C}_3$ is preserved under evolution and the requirement 
of stationarity of ${\cal C}_4$ only determines a restriction on one of the Lagrange 
multipliers, namely, $\lambda_2 = N^3 \Omega / a^2 \Phi$. Thus, we are dealing 
with a wholly constrained theory with first- and second-class constraints, which
is a consequence of the rich gauge symmetry of the RT model. The distintive feature 
of the constraints (\ref{eq:c1}) and (\ref{eq:c2}) instead of $\phi_\mu$ is that 
${\cal C}_1$ and ${\cal C}_2$ are constraints that naturally generate the relationships 
(\ref{eq:C3}) and (\ref{eq:C4}) as befit a higher-order derivative brane theory 
\cite{hambranes}.

Following Dirac's program, the set of constraints should be separated into 
subsets of first- and second-class constraints~\cite{HT}. It is quite well known 
that for each pair of second-class constraints there is one degree of freedom 
which is not physically important and has to be removed from the theory, and for 
each first-class constraint one degree of freedom is removed. For our system we 
have two first-class phase space constraints
\begin{eqnarray}
{\cal F}_1 & = &  {\cal C}_1 \,, 
\label{eq:f1} 
\\
{\cal F}_2 & = &  \frac{N^3 \Omega}{a^2 \Phi} {\cal C}_2 + {\cal C}_3\,,
\label{eq:f2}
\end{eqnarray}
and two second-class constraints. The selection of the second-class constraints 
is straightforward (see the Appendix). We choose them as
\begin{eqnarray}
{\cal S}_1 & = & {\cal C}_2 \,, 
\label{eq:s1} \\
{\cal S}_2 & = & {\cal C}_4 \,.
\label{eq:s2}
\end{eqnarray}
Note that as we have two linear independent first-class constraints, we have the 
presence of two gauge transformations in the RT model. In the Appendix 
we discuss more thoroughly the Poisson brackets among the phase space constraints. 
The counting of degrees of freedom is as follows: $dof = [(\mbox{Total}\ \mathrm{number}\ 
\mathrm{of}\ \mathrm{canonical}\ \mathrm{variables}) 
- 2 \times (\mathrm{first-class}\ 
\mathrm{constraints})~-~(\mathrm{number}\ \mathrm{of}\ \mathrm{second-class}
\newline
 \mathrm{constraints})]/2 = 
[8 - (2 \times 2) - 2]/2 = 1$, which agrees with the number of normals to the
worldvolume \cite{hambranes}. Such a single degree of freedom can be identified as the scale 
factor $a(\tau)$.

\section{Gauge--fixing}

According to the conventional Dirac scheme, in order to extract the physical 
meaningful phase space for a constrained system we need a gauge--fixing prescription 
which entails the introduction of extra constraints, avoiding in this way the 
gauge freedom generated by constraints (\ref{eq:f1}) and (\ref{eq:f2}). To achieve
this we will consider the conventional cosmic gauge condition
\begin{equation}
\label{eq:varphi1}
 \varphi_1 = N - 1= \sqrt{\dot{t}^2 - \dot{a}^2} - 1 \approx 0\,,
\end{equation}
and the generalized evolution Eq. (\ref{eq:master})
\begin{equation}
\label{eq:varphi2}
\varphi_2 =  N^2 + \dot{a}^2 - \gamma\,N^2 H^2 a^2 \approx 0\,,
\end{equation}
where the $\approx$ symbol stands for weak equality in the Dirac approach for
constrained systems \cite{HT}.
From the geometric point of view, this set of gauge conditions is good enough 
since the matrix $\left(\left\{ {\cal F} , \varphi_{1,2} \right\}\right)$ is 
nondegenerate in the constraint surface. Indeed, under the Poisson bracket 
structure~(\ref{eq:PB}), it is straightforward to show that gauges $\varphi_1$ 
and $\varphi_2$ form a second-class algebra with the constraints ${\cal F}_1$ 
and ${\cal F}_2$
\begin{equation}
\begin{split}
\left\lbrace \varphi_1, {\cal F}_1 \right\rbrace &= \varphi_1 + 1, \\
 \left\lbrace \varphi_1, {\cal F}_2 \right\rbrace &=0, \\
\left\lbrace \varphi_2, {\cal F}_1 \right\rbrace &= 2\varphi_2 + 2\gamma H^2 a^2\,, \\
 \left\lbrace \varphi_2, {\cal F}_2 \right\rbrace &= F(a,\dot{a},\dot{t})\,,
\end{split}
\end{equation}
where $ F(a,\dot{a},\dot{t})$ is a complicated function \footnote{This function is given by 
the awkward expression $F= -(2\dot{a}/a) \left[ \dot{t}^2 \left( \frac{\Theta}{\Phi} +2 
\right)  + H^2 \left( \gamma a^2 - \frac{4\Omega^2}{H^6 a^6 (\gamma -1)(3\gamma -1)}\right)  
\right] $.}. Consequently, velocities $\dot{t}$ and $\dot{a}$ must be discarded as dynamical degrees 
of freedom.

The use of the completeness relation (\ref{eq:Hmunu}) results 
efficient at this level: It allows us to express the quantity $P^2$  
by the equivalent expression $- (P\cdot \eta)^2 + (P\cdot n)^2$. This suggests the implementation 
of the following canonical transformation to a new set of phase space 
variables:
\begin{equation}
\begin{split}
N     & :=  \sqrt{\dot{t}^2 - \dot{a}^2}\,,   \\
\Pi_N & :=  \frac{1}{N}(P\cdot \dot{X}) \,,   \\
v     & :=  \mathrm{arctanh} \left( -\frac{p_a}{p_t}\right) \,,\\
\Pi_v & :=  N (P\cdot n) \,,
\end{split} 
\label{eq:transf}
\end{equation}
together with the coordinate transformation $Z^\mu := X^\mu - \left\lbrace 
X^\mu , v \right\rbrace \,\Pi_v $, while the momenta $p_\mu$ remains unaltered. 
Such transformation can be physically interpreted as a Lorentz rotation in 
phase space which, straightforwardly, preserves the structure of the canonical 
Poisson brackets
\begin{equation}
\begin{split}
 \left\lbrace N , \Pi_N \right\rbrace &=  1= \left\lbrace v , \Pi_v \right\rbrace, \\
 \left\lbrace Z^\mu  , p_\nu \right\rbrace &= \delta^\mu {}_\nu\,,
\end{split}
\end{equation}
as expected. A transformation of the same kind was considered in \cite{kapu} for a quantum
treatment of a kink model. To understand in a more transparent way the geometric content 
of this transformation let us go through the intrinsic angular momentum density 
of the RT model calculation carefully. In terms of the new phase space variables 
the intrinsic angular momentum is given by $ {\cal M}^{\mu \nu} _i  = 
\eta_a {\cal M}^{a \,\mu \nu} _i =  N P^{[\mu} \dot{X}^{\nu ]}$~\cite{noether}.
Such momentum is conserved in the sense that $\nabla_a {\cal M}^{a \,\mu \nu} _i = 0$.
This contribution shows up as an effect of the finite width of the brane in
comparison with flimsy branes described by Dirac-Nambu-Goto action. The only nonvanishing 
contribution to the intrinsic angular momentum reads
\begin{equation}
{\cal M}_{i \,\,\,\eta n} = \frac{1}{2}N \Pi_v \,,
\end{equation}
which in turn tell us that our new variable $\Pi_v$ can be thought of as 
an angular momentum component.

In attempting to take into account the new phase space variables, the first- and 
second-class constraints (\ref{eq:f1}-\ref{eq:s2}) become
\begin{eqnarray}
\label{eq:F1}
{\cal F}_1 &=& N \Pi_N\,, \\
\label{eq:F2}
{\cal F}_2 &=&  p\cdot \dot{X} + N\left(  a^3 H^2 + \frac{1}{a^3}N^2 \Pi_N ^2
- \frac{1}{a^3}\Pi_v ^2  \right) , 
\end{eqnarray}
and
\begin{eqnarray}
 {\cal S}_1 &=& \Pi_v - \frac{a^2\dot{t}}{N}=0,
\label{eq:SS1} \\
{\cal S}_2 &=& p_a \dot{t} + p_t \dot{a} = \mathfrak{p}_a \dot{t} - \frac{2a\dot{a}}{N}
\dot{t} = 0\,,
\label{eq:SS2}
\end{eqnarray}
respectively.  Note that in the second-class constraint~(\ref{eq:SS2}) 
we split the momentum conjugated to $a$ according to 
relation~(\ref{eq:pa-terms}), and hence, 
the second-class constraint (\ref{eq:SS2}) 
results in identity~(\ref{eq:boldq}).
Furthermore, second-class identities~(\ref{eq:SS1}) and~(\ref{eq:SS2}) 
will become auspicious at the quantum level since they
enclose important operator identities. 

One can develop further the constraint  ${\cal F}_2$ (\ref{eq:F2}) by expressing the 
velocities in terms of the momenta, $\dot{a} = -(N/a)\left[ \mathbf{p}_a 
/ \left((\gamma -1)a^2 H^2 + 2 \right)\right] $ and $\dot{t} = - \Omega N 
/ (\gamma -1) a^3 H^2$ by using the second gauge~(\ref{eq:varphi2}). 
Thus, a direct calculation on the first term in~(\ref{eq:F2}) yields
\beq
 p\cdot \dot{X} & = & 
-\left( \frac{N}{a[(\gamma - 1)a^2H^2 + 2]} \right)\mathbf{p}_a ^2 \nn\\
& & -
\left(\frac{N\Omega}{(\gamma - 1)a^3H^2}\right) p_t  + \mathfrak{p}_a \dot{a}.
\eeq
Finally, after a lenghty but straightforward computation
the constraint ${\cal F}_2$ can be cast as
\begin{widetext}
 \begin{equation}
{\cal F}_2 =  N\left\{
\mathbf{p}_a ^2 - a \left[ -
\left(\frac{\Omega}{(\gamma - 1)a^3H^2}\right) p_t  + \frac{\mathfrak{p}_a \dot{a}}{N} 
 +  a^3 H^2 + \frac{1}{a^3}N^2 \Pi_N ^2
- \frac{1}{a^3}\Pi_v ^2   \right]  \left[\ (\gamma -1)a^2 H^2 + 2  \right] \right\} .
\label{eq:quantize}
\end{equation}
\end{widetext}
We could also use Eq. ~(\ref{eq:pt}) and the second-class 
condition~(\ref{eq:SS2}), which impose the identities $p_t=-\Omega$
and $\mathfrak{p}_a=2a\dot{a}/N$, respectively, to conclude that the constraint 
${\cal F}_2 \approx 0$ gives rise to a quadratic expression for the involved momenta.
That is, the second gauge condition (\ref{eq:varphi2}) shifts the problem from the linear
dependence in the momenta, to a convenient quadratic expression for the physical 
momenta. To close this section we must mention that the constraints ${\cal F}_1$ and 
${\cal F}_2$ form an algebra isomorphic to the Lie algebra associated to the lower 
triangular subgroup of $SL(2,\mathbb{R})$ as argued in the Appendix.
This is the starting point to achieve an algebraic quantization as we will sketch
below.

\section{Quantization}
\label{sec:quantum}

In this section we study the canonical quantization of our system. Also,  
we will sketch an alternative different quantum theory for our model which emerge 
from the corresponding first-class symmetries. To this end, we emphasize the totally 
dissimilar nature which first- and second-class constraints play in the quantum 
theory, and also, we explore the different senses in which the physical states 
of our theory can be defined.

We start in the conventional way by promoting 
the classical constraints into operators, densely defined on
a common domain in a proper Hilbert space.
As it is well known, we can only achieve a  consistent classical theory 
by implementation of the Dirac bracket.  Once this is done, the 
second-class constraints are eliminated off the theory by converting 
them into strong identities.  
At the quantum level this is mirrored by defining the 
quantum commutator of two quantum operators as 
\beq
\label{eq:commutator}
[\hat{A},\hat{B}]:=i\widehat{\lbrace A,B \rbrace^\ast} \,,
\eeq
where the Dirac bracket $\lbrace\cdot,\cdot\rbrace^\ast$ is defined in
Eq.~(\ref{eq:DB}).  Thus, with this prescription the operators 
corresponding to second-class constraints are also enforced as operator 
identities~\cite{HT}.  For our system, this yields the quantum operator 
expressions
\beq
\label{eq:hatS1}
\hat{{\cal S}}_1 &=& \hat{\Pi}_v - \widehat{\frac{a^2\dot{t}}{N}}=0 \,, \\
\label{eq:hatS2}
\hat{{\cal S}}_2 &=& \hat{\mathfrak{p}}_a  - \widehat{\frac{2a\dot{a}}{N}} = 0  \,,
\eeq
which, in particular, tell us the character of the quantum operators
$\hat{\Pi}_v$ and $\hat{\mathfrak{p}}_a$. For the rest of the variables, we choose 
to work on the ``position'' representation, where we consider the position 
operators by multiplication and their associated momenta operators by $-i$ times 
the corresponding derivative operator when applied on states defined on a 
suitable Hilbert space.

By defining the quantum first-class constraints as
\begin{widetext}
\beq
\label{eq:quantumF1}
\hat{{\cal F}}_1 &:=& -iN\frac{\partial\ }{\partial N}    \,,\\
\label{eq:quantumF2}
\hat{{\cal F}}_2 &:=& N\left\{ -\frac{\partial^2\ }{\partial a^2} 
- \left[\frac{i\Omega}{(\gamma-1)a^3H^2} \frac{\partial\ }{\partial t} 
+ 2a(\gamma H^2 a^2 - 1) + a(1 - \gamma)H^2 a^2 
- \frac{1}{a^3}\left( N\frac{\partial\ }{\partial N} \right)^2
\right] \right. \nn \\ &&\quad \left. \times a\left[ (\gamma -1)H^2a^2  + 2  \right]
\right\}   \,,
\eeq
\end{widetext}
we will work on the assumption that the commutators of these quantum 
constraints form a closed Lie algebra which will be also isomorphic to 
the algebra $\mathfrak{g}$. In fact, the classical first-class constraints
are isomorphic to the algebra $\mathfrak{g}$ associated to the lower triangular 
subgroup $G$ of $SL(2,\mathbb{R})$ (see the Appendix).
Quantization of the lower triangular subgroup of 
$SL(2,\mathbb{R})$ by algebraic methods 
was extensively studied in~\cite{LouMol2} 
(see also~\cite{LouMol1} for comparison).  
Now we explore the rather different senses in which the 
quantum constraints can be used to define appropriate physical states.

\subsection{Na\"ive Dirac constraints}
\label{sec:Dirac}

\begin{figure}
  \caption{WDW potential for geodetic gravity with na\"ive Dirac constraints.}
  \begin{center}
    \includegraphics[angle=0,width=8cm,height=4cm]{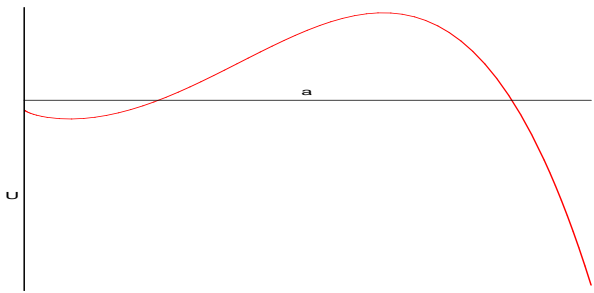}
    \label{fig:Davidson}
  \end{center}
 \end{figure}

First, we explore the Wheeler--DeWitt 
equation emerging by considering the physical states $\Psi$ 
of the theory as those defined by na\"ive Dirac conditions
\beq
\label{eq:Dirac1}
\hat{{\cal F}}_1 \Psi & = &   0   \,,\\  
\label{eq:Dirac2}
\hat{{\cal F}}_2 \Psi & = &   0   \,.
\eeq
Equation (\ref{eq:Dirac1}) simply tells us that our physical states $\Psi$ 
are not explicitly depending on the phase space variable $N$.  
However, due mainly to the complexity of our WDW
 Eq. (\ref{eq:Dirac2}), we have not succeed in finding 
explicit solutions for the physically admisible quantum states.
We note that the last term in the operator~(\ref{eq:quantumF2})
will bring a vanishing contribution to the WDW
equation, and also we see that the $t$--dependence 
can be avoided by assuming $\Psi(a,t):=e^{-i\Omega t}\psi$,
where $\psi:=\psi(a)$ satisfies the WDW equation
\beq
\label{eq:WdW}
\left[ - \frac{\partial^2 }{\partial a^2} + U(a) 
 \right] \psi(a) = 0\,,
\eeq 
where the potential $U(a)$ is given by
\begin{equation}
 U(a) = a^2 \left[ (\gamma - 1)H^2 a^2 + 2 \right]^2 \left( 
1 - \gamma H^2 a^2 \right)\,,
\end{equation}
which is recognized as the potential function found in~\cite{DKL} by repeated use of 
the master evolution constraint~(\ref{eq:master}) in Eq.~(\ref{eq:Dirac2}). 
The behavior of this potential is drawn in Fig.~\ref{fig:Davidson},
where we can see the characteristic potential barrier.  
As discussed by Davidson and coworkers, it can be shown that after 
considering appropriate boundary conditions the big--bang singularity 
in our quantum theory can be neutralized by properly choosing the origin 
as inaccessible to wave packets. For further details on the behavior of 
the potential $U(a)$, the reader is referred to~\cite{DKL}.

\subsection{Modified Dirac constraints}
\label{sec:modDirac}

As discussed in Refs.~\cite{GM1,tuynman}, there exists a procedure
which allows us to reduce nonunimodular groups to unimodular ones 
and this in turn brings a remarkable alteration for systems amenable to geometric or 
algebraic quantization which comprises a modification for the 
Dirac conditions on physical states.  Let $\{ \hat{C}_a \}$
be a set of quantum constraints operators that generate a 
nonunimodular gauge group with the commutators $[\hat{C}_a,\hat{C}_b]=if^c_{ab}\hat{C}_c$,
where $f^a_{bc}$ are the structure constants of the corresponding Lie algebra.
Thus, the ``unimodularization'' procedure for nonunimodular groups dictates the consideration
of the physical states $|\Psi\rangle$ as those satisfying $\hat{C}_a|\Psi\rangle=-(i/2)f^b_{ab}|\Psi\rangle$.  
Such modified Dirac conditions agree with the na\"ive Dirac constraints if, and only if,
the group is unimodular. 

Accordingly, for our theory the 
modified Dirac conditions 
for the gauge group invariant
quantization of the system can be shown to be equivalent to 
\beq
\label{eq:qstates1}
\left[\hat{{\cal F}}_1 -\frac{i}{2} \right]|\Psi\rangle & = &  0   \,,\\
\label{eq:qstates2}
\hat{{\cal F}}_2 |\Psi\rangle                           & = &  0   \,,
\eeq
which consequently define physical states $|\Psi\rangle$. 
Equation (\ref{eq:qstates1}) is 
equivalent to the homogeneity condition 
$|\Psi(rN)\rangle=r^{-1/2}|\Psi(N)\rangle$ for $r>0$ \cite{LouMol2}.  
Further, (\ref{eq:qstates1}) can be explicity solved
by taking $|\Psi\rangle=\frac{A}{N^{1/2}}|\psi\rangle$, 
where $A$ is a constant, and $|\psi\rangle$ is a function of the variables $a$ and $t$.
Once more we also do not have control on the 
explicit solutions for the physically admisible quantum states.  
The $t$--dependence 
can be avoided by assuming $|\psi(a,t)\rangle:=e^{-i\Omega t}|\varphi\rangle$,
where $|\varphi\rangle$ is thought of as a function of the scale factor $a$ only,
which satisfies the WDW equation
\beq
\label{eq:WdW-1}
\left[ - \frac{\partial^2 }{\partial a^2} + U(a) 
+ \frac{(\gamma -1)a^2 H^2 + 2}{4a^2} \right] |\varphi(a)\rangle = 0\,,
\eeq 
where the potential $U(a)$ was described in the previous subsection.
Hence, we see that our modified quantum theory brings out an extra 
potential term into our WDW equation, which succinctly differs from the one 
found with the na\"ive Dirac procedure.

\begin{figure}
  \caption{WDW potential for geodetic gravity with modified Dirac constraints.}
  \begin{center}
    \includegraphics[angle=0,width=8cm,height=4cm]{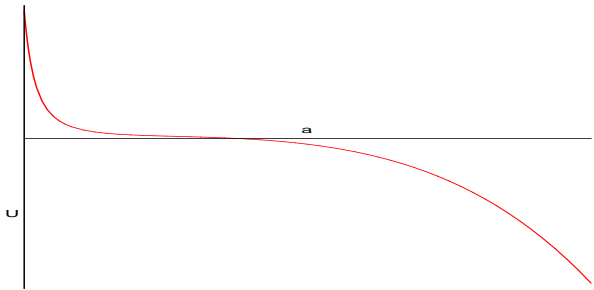}
    \label{fig:refinedplot3}
  \end{center}
 \end{figure}

We note that the extra term is purely emerging from the modified quantum 
Dirac Eqs.~(\ref{eq:qstates1}) and~(\ref{eq:qstates2}), and it is 
completely absent while considering the na\"ive Dirac procedure. This term 
will be nonvanishing even in the Einstein limit $(\gamma\rightarrow 1)$, 
where it goes as $a^{-2}$.  Further studies about the possible physical 
implications of this term could be carried out. The behavior of the modified 
potential is drawn in Fig.~\ref{fig:refinedplot3}, where we can notably see 
that the central barrier potential present in Fig.~\ref{fig:Davidson} is 
almost vanishing while an infinite barrier emerges at the origin.
Until our knowledge, the resulting unbounded potential is not realistic
despite the first-class constraints suggest this modified description.
Nevertheless, one can not resist the speculation of such possible quantum behavior.
Thus, rather than a nice potential, this time it is a more complicated function
with distinct features notwithstanding the internal constraint symmetries that demand an 
unimodularization procedure.

\section{Concluding remarks}

By making use of the Ostrogradski formalism we have developed an alternative 
Hamiltonian description of the RT brane model. 
Unlike the Hamiltonian treatment by Davidson and coworkers for this model 
\cite{davidson0,davidson1}, our analysis above keeps the original variables without the
necessity of introducing nondynamical variables. 
At first sight, this may look like an unnecessary complication since the configuration space
is initially increased only to be reduced again at a later stage by
imposing the constraints and fixing the gauge. Nevertheless, it is hoped
that despite computational complications we have provided an improvement
of physical clarity, in particular, concerning the geometrical meaning of the
constraints of the theory and the physical content of the achieved
canonical transformation. Although the 
Ostrogradski approach has a price to pay, since
neither the momentum $P$ has the meaning of mechanical momentum nor $H_0$
has to do with the energy of the system, as it is customary, 
these quantities are adequate for providing a set of canonical equations which correctly 
describe the evolution of the system.  
Also, an important point to mention is that the formalism is rich enough to
demonstrate the real role of  the  
extra terms coming from the surface:  the phase space constraints 
of the system impose identities 
for these quantities which are valid at both classical and quantum levels, hence
eliminating the unphysical degrees of freedom. 

In spite of the fact that this model is a 
second-order derivative theory and since it is well known that at quantum level 
the energy for higher-order derivative Lagrangians is almost always unbounded below, 
the RT model results an exception due to its linear dependence on the accelerations
which in turn contain important physical information commonly absent in higher-order
derivative theories~\cite{simon}. Like it or not, until now a Hamiltonian
approach for the RT field theory demands the use of extra unphysical degrees of freedom
at the beginning which by means of the phase space constraints are frozen out.
Our adopted treatment renders the passage to a full quantization
for the system which can be achieved by means of an inspired canonical transformation. 
We conclude further that the Ostrogradski quantum approach has exactly the same unique degree of freedom
as the Davidson and coworkers approach.  
Although our Wheeler-DeWitt equation for the scale factor is not analytically 
manageable, it is good enough to
substract from it some interesting features.
In particular, the potential we found is exactly the same as 
the one extensively discussed by Davidson et al.~\cite{DKL}. 
Furthermore, our Hamiltonian approach makes feasible the
quantum treatment of Lagrangians with linear higher-order
derivative dependence in the fields.

It is suggesting that relativistic theories linear in the accelerations, 
for which characteristic surface terms are commonly neglected, 
are, as a generic feature, reluctant to quantization.
To present day, quantization for these kind of systems 
have been mainly studied by considering some extra degrees 
of freedom by several other methods.  From this point of view, 
our intention has also been to introduce the Hamiltonian-Ostrogradski
approach as a geometrical powerful method to beset this sort of
systems. An specific example of this would be to apply 
our treatment to the almost forgotten idea concerning the rigid bubble electron,
for which a linear
correction in the extrinsic curvature of the electron surface is added
to the Dirac--Nambu--Goto action in contrast to the conventional 
first order method where a surface
term is omitted~\cite{electron}. It will be worked elsewhere.

\acknowledgments
This work was
partially supported by SNI (Mexico).  
ER and RC acknowledge support from CONACYT (Mexico) 
research grant J1-60621-I. RC also acknowledge 
support from EDI, COFAA--IPN and SIP-20082476.  AM was supported 
by CONACYT (Mexico) grant CB-75629. ER acknowledge
partial support from PROMEP, 2006-2008 (Mexico).

\appendix
\section{Algebraic properties of the constraints}
\label{sec:second}

We can construct the matrix ${\cal C}_{AB}$ whose elements are the Poisson 
brackets of all the constraints ${\cal C}_A$ where $A,B=1,2,3,4$. Hence,
\begin{equation}
({\cal C}_{AB})= \frac{1}{aN}\left(
\begin{array}{cccc}
0&0&0 & 0 \\
0&0&0& -a^2 \Phi \\
0&0&0& N^3 \Omega \\ 
0& a^2 \Phi & -N^3 \Omega &0
\end{array}
\right)\,,
\label{eq:matrix}
\end{equation}
in the constraint surface.
This matrix has rank 2, which is a signal that we have two first-class 
constraints~\cite{HT}. It is also important to mention that 
constraints ${\cal C}_1$ and ${\cal C}_3$
form an algebra, namely,
\begin{equation}
\begin{split}
\left\lbrace {\cal C}_1, {\cal C}_1 \right\rbrace &= 0, \\ 
\left\lbrace {\cal C}_1, {\cal C}_3 \right\rbrace &= - {\cal C}_3\,, \\
\left\lbrace {\cal C}_3, {\cal C}_3 \right\rbrace &= 0\,,
\end{split}
\end{equation}
which reflects the invariance under reparametrizations of the RT field theory
as a fundamental gauge symmetry.  Indeed, this algebra 
results an isomorphism of the Lie algebra $\mathfrak{g}$ associated to 
the lower triangular subgroup of $SL(2,\mathbb{R})$ with positive 
diagonal elements, $G$.  Such Lie algebra $\mathfrak{g}$ is 
spanned by the matrices~\cite{Howe}
\begin{equation}
h :=
\left(
\begin{array}{cc}
1 & 0
\\
0 & -1
\end{array}
\right)
\ \ , \ \
e^- :=
\left(
\begin{array}{cc}
0 & 0
\\
1 & 0
\end{array}
\right)
\ \ ,
\end{equation}
whose commutator is
\begin{equation}
\left[ h \, , \,  e^-  \right]
=
-2e^-
\,,
\label{eq:g-basiscomm}
\end{equation} 
and hence we realize the isomorphism through the identification
${\cal C}_1\mapsto h/2$ and ${\cal C}_3 \mapsto e^-$.  
Among the relevant properties of the subgroup $G$ we refer
that $G$ is two--dimensional, non--Abelian, connected, and nonunimodular.
This last property plays an important role in our quantum 
theory, as developed in section~\ref{sec:quantum}.

Also, among the second-class constraints, (\ref{eq:s1}) and 
(\ref{eq:s2}), we can construct the matrix ${\cal S}_{IJ} = \left\lbrace 
{\cal S}_I , {\cal S}_J \right\rbrace $, given by
\begin{equation}
 ({\cal S}_{IJ}) = \frac{a\Phi}{N}  
\left( 
\begin{array}{cc}
0 & -1\\
1 & 0
\end{array}
\right) \,,
\end{equation}
and its inverse 
\begin{equation}
 ({\cal S}^{IJ}) = \frac{N}{a\Phi}  
\left( 
\begin{array}{cc}
0 & 1\\
-1 & 0
\end{array}
\right) \,,
\end{equation}
where $I,J = 1,2$.   The matrix ${\cal S}^{IJ}$ will
help us to construct a Dirac bracket in the standard way:
Let $f$ and $g$ be two arbitrary functions then
\beq
\label{eq:DB}
\lbrace f,g \rbrace^{\ast}:=\lbrace f,g\rbrace -
\sum \lbrace f,{\cal S}_I\rbrace {\cal S}^{IJ} \lbrace {\cal S}_J,g\rbrace \,,
\eeq
where $\lbrace\cdot,\cdot\rbrace$ stands for the Poisson bracket defined 
in~(\ref{eq:PB}).  As it is well known, classically, the Dirac bracket 
is essential to eliminate the second-class constraint off the theory
by converting  them into simple functional identities.  
The need for the Dirac bracket 
is also very relevant
at the quantum theory since we can only reach a consistent quantization procedure
through the implementation of this bracket~\cite{HT}.

\newpage 


\end{document}